\documentclass[12pt]{aastex}

\shorttitle{Nearby Dwarf}
\shortauthors{Massey, Henning, \& 
Kraan-Korteweg}

\begin{document}

\title{A Neighboring Dwarf Irregular Galaxy
Hidden by the Milky Way} 

\author{Philip Massey\altaffilmark{1}}

\affil{Lowell Observatory, 1400 W. Mars Hill Road, Flagstaff, AZ 86001}
\email{Phil.Massey@lowell.edu}

\author{P. A. Henning}

\affil{Institute for Astrophysics, University of New Mexico,
800 Yale Boulevard, NE, Albuquerque, NM 87131-1156}
\email{henning@as.unm.edu}

\and

\author{R. C. Kraan-Korteweg}

\affil{Depto.~de Astronom\'{i}a, 
Universidad de Guanajuato, Apartado Postal 144,\\ 
Guanajuato, Gto 36000, M\'{e}xico}
\email{kraan@astro.ugto.mx}

\altaffiltext{1}{
Visiting astronomer, Kitt Peak National Observatory,
a division of the National Optical Astronomy Observatory, which is
operated by the Association of Universities for Research in Astronomy,
Inc., under cooperative agreement with the National Science Foundation.}

\begin{abstract}

We have obtained VLA and optical follow-up observations of the
low-velocity H~I source HIZSS~3 discovered by Henning et al.~2000
and Rivers 2000 in a survey for nearby galaxies hidden by the disk 
of the Milky Way.
Its radio characteristics are consistent with this being a nearby
($\sim 1.8$~Mpc) low-mass dwarf irregular galaxy (dIm).  
Our optical imaging failed to reveal a resolved stellar population, but
did detect an extended H$\alpha$ emission region.  The location of the
H$\alpha$ source is coincident with a partially-resolved H~I cloud in the 
21-cm map. Spectroscopy confirms
that the H$\alpha$ source has a similar radial velocity 
to that of the H~I emission at this location,
and thus we have identified an optical counterpart.  The H$\alpha$
emission (100~pc in diameter 
and with a luminosity of 
$1.4\times10^{38}$ ergs s$^{-1}$) is characteristic of a single H~II region
containing a modest population of OB stars.
The galaxy's 
radial velocity and distance from
the solar apex suggests that it is not a Local Group member, although
a more accurate distance is needed to be certain.
The properties of HIZSS~3 are comparable to those of GR~8, a nearby dIm
with a modest amount of current star formation. Further
observations are needed to characterize its stellar population, determine
the chemical abundances, and obtain a more reliable distance estimate.

\end{abstract}


\section{Introduction}

A complete census of nearby galaxies is hampered by the presence of
our own Milky Way, whose dust and gas creates a ``zone of avoidance"
(ZOA) in the distribution of galaxies on the plane of the sky.  Yet an
accurate knowledge of the mass distribution within our neighborhood
is essential if we are to understand the dynamical evolution of the
Local Group from kinematic studies (e.g., Peebles et al.\ 2001).
In addition, the discovery of previously unknown nearby galaxies will
further efforts to understand the local velocity field (see
Kraan-Korteweg 1986 and Karachentsev et al.~2002) 
as well as providing
additional examples for studying the resolved stellar content of nearby
systems (cf. Mateo 1998, Massey 2003).

Kerr \& Henning (1987) demonstrated the power of single-dish H~I surveys
in searching for hidden galaxies.
Such observations provide not only the two-dimensional location of
galaxies on the plane of the sky, but also the redshift.  Furthermore,
the width of the 21-cm line gives some indication of the mass. 
``Shallow" surveys are now complete in both the north (Henning et al.\ 1998, 
Rivers, Henning, \& Kraan-Korteweg 1999, Rivers 2000) and the 
south (Henning et al.\ 2000),
with more sensitive searches in progress. 
Although no massive Local Group galaxies
have been found lurking behind the Galactic disk, 
the northern survey did reveal a  previously unknown 
spiral galaxy (``Dwingeloo 1") at 3~Mpc (Kraan-Korteweg et al.\ 1994), 
a likely member of the IC~342/Maffei Group.

Here we report on follow-up observations of a galaxy that is nearly
half that distance.  In their survey of H~I sources in the southern ZOA,
Henning et al.\ (2000) lists object HIZSS~3  as having  a heliocentric
radial velocity of only 299 km~s$^{-1}$.  Improved data (presented below)
yield a velocity of 280~km~s$^{-1}$.  Given its location relative to
the solar apex, this translates to a velocity with respect to the Local
Group of 134 km~s$^{-1}$ (following Courteau \& van den Bergh 1999)
implying a distance of only 1.8~Mpc ($H_o$=75 km~s$^{-1}$~Mpc$^{-1}$)
in the absence of any peculiar velocities.  Thus, this is one of the
nearest galaxies known, similar in distance to the well-known Sextans A,
Sextans B, and GR~8 galaxies, nearby but slightly beyond the zero-velocity
envelope that defines the Local Group (Mateo 1998, van den Bergh 2000).
Its location less than 0.1 degrees from the Galactic plane has made
prior optical identification unlikely; nothing but foreground stars is
seen on the Palomar Sky Survey prints or in
the 2MASS images at this position.  There is no object listed within 
5 arcmin of this position in the
Extended Source Catalog (Jarrett et al.~2000) of the 2MASS survey.

The properties of this object (summarized in Table~1) are strongly
suggestive of a dwarf irregular (dIm) galaxy.  The H~I velocity
width at 20\% peak intensity determined by Henning et al.~(2000) is 85
km~s$^{-1}$, a value which is typical for dwarf irregulars (Hunter 1997;
see her Fig.~5).  In addition, new observations (described below) yield
an H~I diameter of 6 arcminutes.  At a distance of 1.8~Mpc, this would
correspond to a diameter of 3~kpc, a value which is also typical for
dwarf irregulars (Hunter 1997, Fig.~4).

\section{New Observations}

\subsection{VLA}

The galaxy was originally discovered at 21-cm by the Dwingeloo Obscured
Galaxies Survey of the northern ZOA ($30^\circ \le \ell \le 220^\circ$,
$|b|$~$\leq$~5$^\circ$; Henning et al.\ 1998, 
Rivers, Henning, \& Kraan-Korteweg
1999, Rivers 2000), and it was also detected by the H~I Parkes Zone
of Avoidance Shallow Survey of the southern ZOA 
($212^\circ \le \ell \le 36^\circ$,
$|b|$~$\leq$~5$^\circ$; 
Henning et al.\ 2000).
The galaxy, at galactic coordinates $\ell = 218^\circ$, $b = 0.1^\circ$ 
lies in a region of overlap between the two surveys.
The basic H~I parameters were first published
as part of the latter catalog, hence the HIZSS moniker.

H~I synthesis data were obtained with the VLA\footnote{The VLA is
operated by the National Radio Astronomy Observatory, a facility of
the National Science Foundation operated under cooperative agreement
by Associated Universities, Inc.} in 1999 as part of
follow-up observations for the Dwingeloo project (Rivers 2000).
A 30-minute observation was made in the compact D-array, and the maps were
created using natural weighting of the visibilities.  This produced an angular
resolution, as measured by the half-power major and minor axes of the
synthesized beam, of 66 $\times$ 47 arcsec.  The velocity resolution obtained
was 10 km~s$^{-1}$.  Summing all channels containing H~I emission produced the
H~I total intensity map presented as Fig.~\ref{fig:totalHI}.
Taking the estimated distance of 1.8 Mpc and
the total integrated flux from these interferometric observations,
measured within an H~I column density of 1 M$_\odot$~pc$^{-2}$,
of 24.9 Jy~km~s$^{-1}$, the H~I mass
of this galaxy would be $1.9 \times 10^7$ M$_{\odot}$.  
The flux recovered with the
single-dish Parkes survey was slightly higher: 32.1~Jy~km~s$^{-1}$
(Henning et al.\ 2000),
thus yielding a somewhat higher estimated H~I mass, 
$2.5 \times 10^7$ M$_{\odot}$.
(The slightly higher value is not unexpected, since the Parkes measurement was
not cut off at the 1 M$_{\odot}$~pc$^{-2}$ column density level.)
More accurate determination of the distance will be required to pin down
its exact mass, although this at the low end of what is observed for
late-type dwarfs (Swaters 1999).
The H~I diameter measured to the 1 M$_{\odot}$~pc$^{-2}$ level is 6 arcmin,
corresponding to 3~kpc.
The velocity field, shown in Fig.~\ref{fig:HIvel}, indicates fairly uniform rotation.
The velocity width measured at 20\% of the peak flux density from these
observations is 91 km~s$^{-1}$, consistent
with the value of 85 km~s$^{-1}$ from the Parkes observations,
and also consistent with HIZSS 3's being a low-mass dIm galaxy.

The peak of the H~I distribution is at 
$\alpha_{\rm J2000}=7^{\rm h}00^{\rm m}29.3^{\rm s}$,
$\delta_{\rm J2000}=-04^\circ12'30"$, with a heliocentric velocity of
280 km~s$^{-1}$.

\subsection{Optical Identification}

We expected difficulties in the optical identification due to the 
low galactic latitude and high foreground extinction: Henning et al.~(2000)
estimate $A_B=4.7$ mags based upon the DIRBE/IRAS extinction maps
(Schlegel, Finkbeiner, \& Davis 1998); i.e., $E(B-V)=1.1$~mags.   
At the time we began
our search  for an optical counterpart (April 2003) the object was 
observable only for the first hour or so of the night.

Direct images were taken centered on the center of the VLA map with
Lowell Observatory's 1.1-m Hall telescope using a SITe 2048$\times$2048
CCD on UT 8 and 9 April 2003.  The CCD was binned $2\times 2$ for a scale
of 1.13 arcsec pixel$^{-1}$.  The field of view was 19.4 arcmin
by 19.4 arcmin.

Broad-band {\it BVRI} images obtained on the first night failed to show
anything other than the expanded swarm of foreground galactic stars.  
However, an H$\alpha$ exposure on
the second night clearly showed an extended object about 70 arcsec from
the peak of the H~I distribution.
In Fig.~\ref{fig:images} we compare the H$\alpha$ image with
an $R$-band
exposure. The H$\alpha$ image (Fig.~\ref{fig:images}a) 
is the sum of four 15-minute integrations,
while the R-band exposure (Fig.~\ref{fig:images}b) 
is the sum of 15 1-minute exposures, taken
in this way to minimize saturation.
The seeing on each frame was approximately 3 arcsecs, due in large
part to the high airmass ($\sim 1.8$) of the observations.  We also show
in Fig.~\ref{fig:images}c the residual image obtained by 
subtracting the scaled R-band exposure from the H$\alpha$
exposure after sky has been removed from each.  

\begin{figure}
\epsscale{0.65}
\plotone{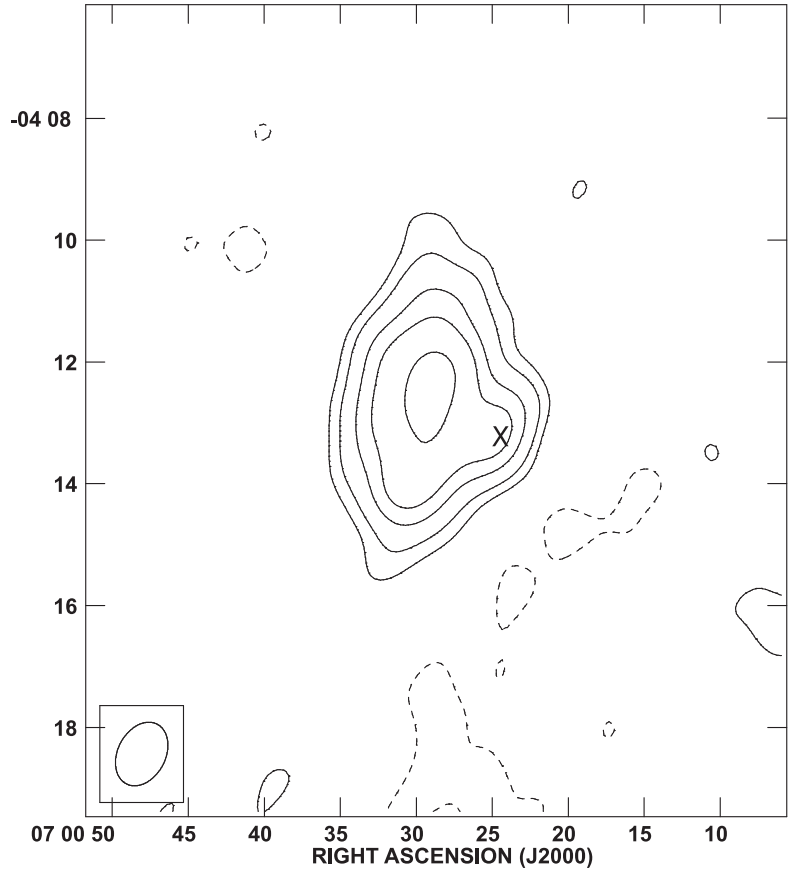}
\caption{\label{fig:totalHI}  H~I total intensity map of HIZSS 3.
Contours are integrated H~I emission with levels at -4, -2, 2, 4, 8, 12, and
20 times the rms noise of $1.6 \times 10^{-2}$ Jy beam$^{-1}$.  The synthesized
beam is represented at the lower-left-hand corner.  The ``X" denotes the
location of the H$\alpha$ source.
}
\end{figure}

\begin{figure}
\epsscale{0.65}
\plotone{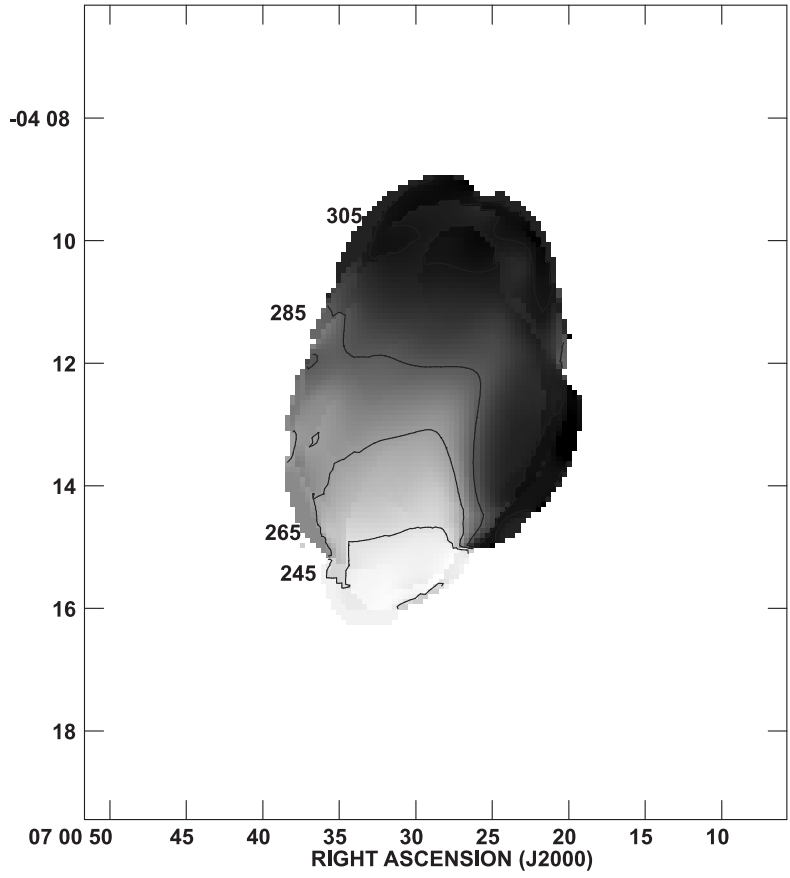}
\caption{\label{fig:HIvel} H~I velocity field of HIZSS 3.
Contours are heliocentric velocity in km~s$^{-1}$, superposed on a
greyscale image of the velocity field map.
}
\end{figure}

\begin{figure}
\epsscale{0.40}
\plotone{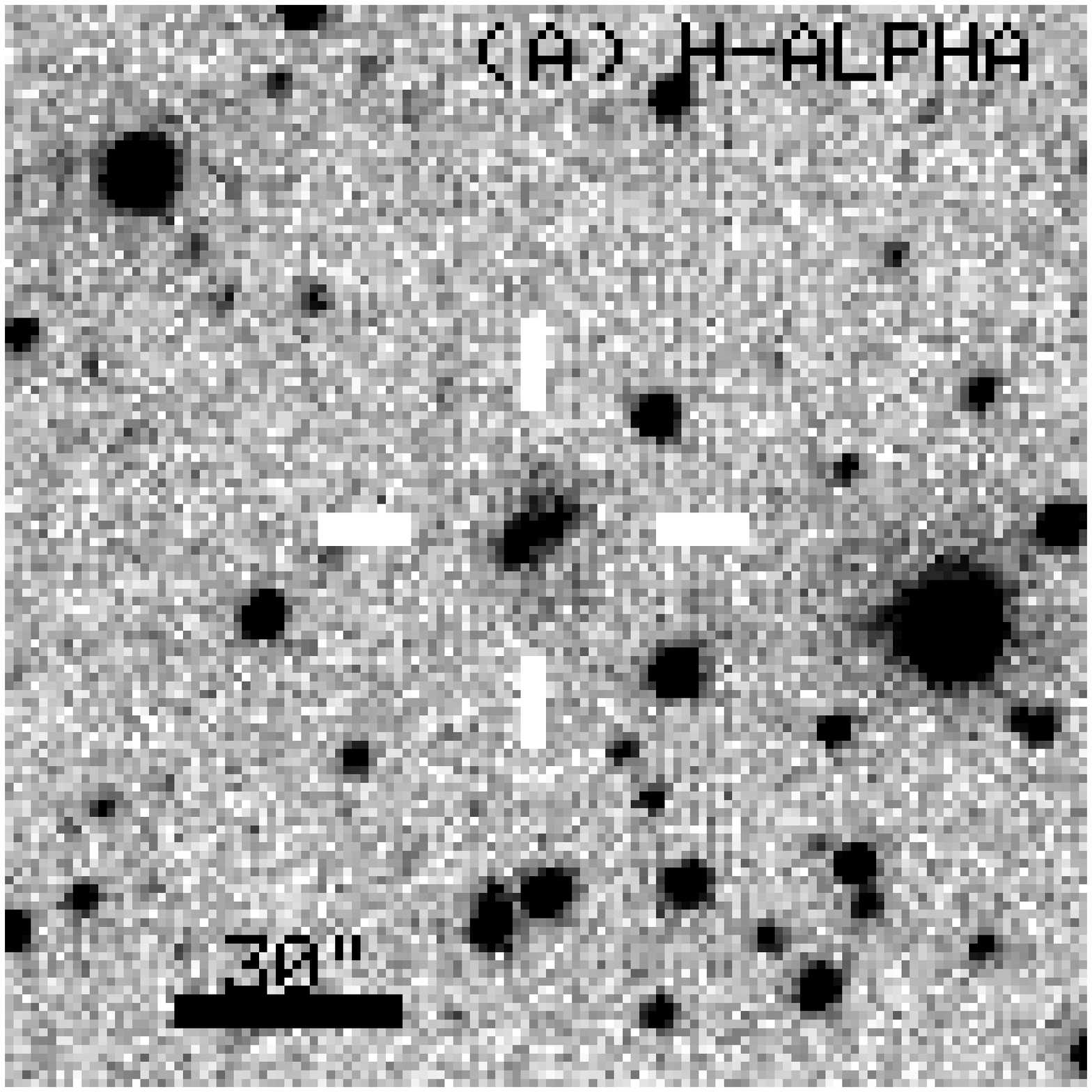}
\plotone{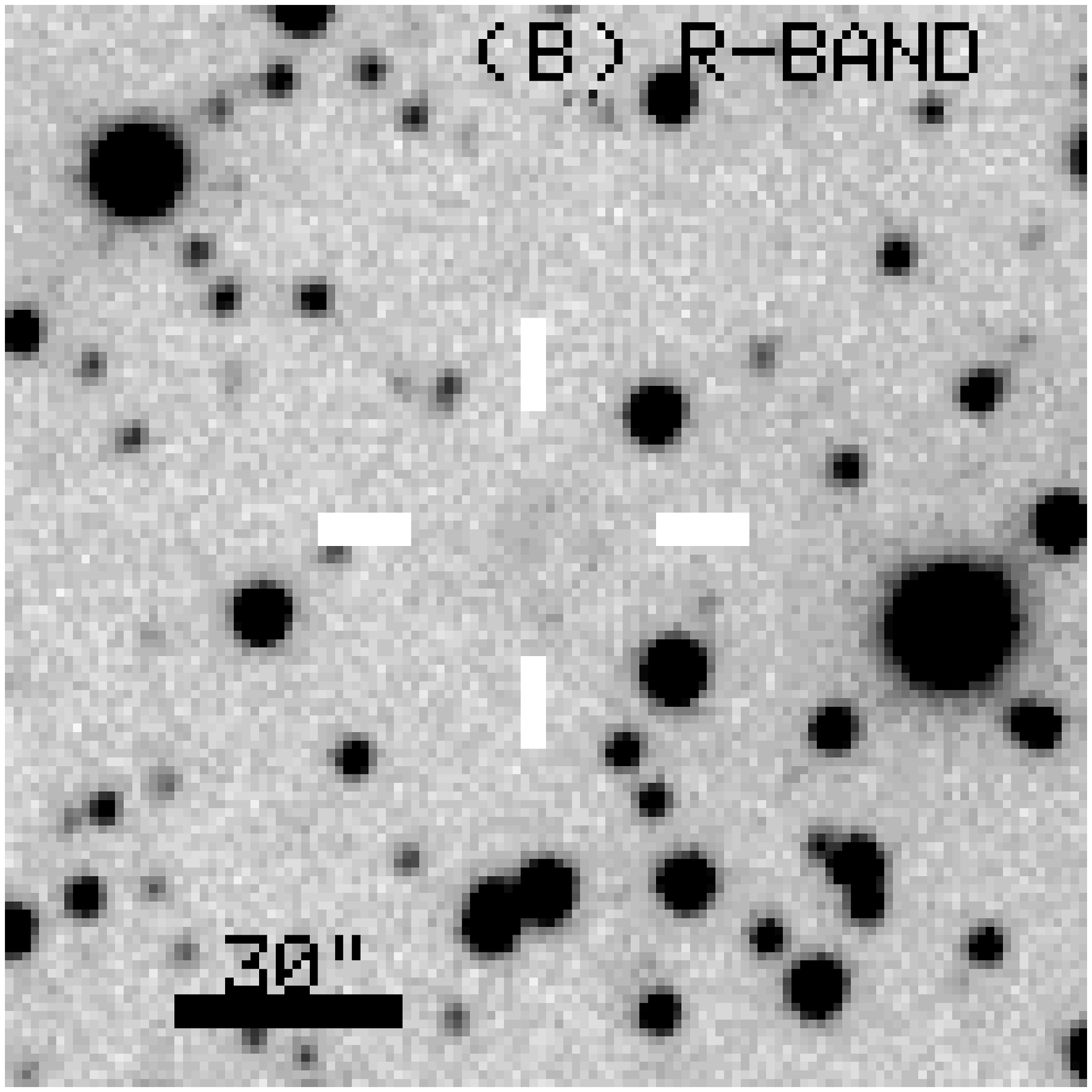}
\plotone{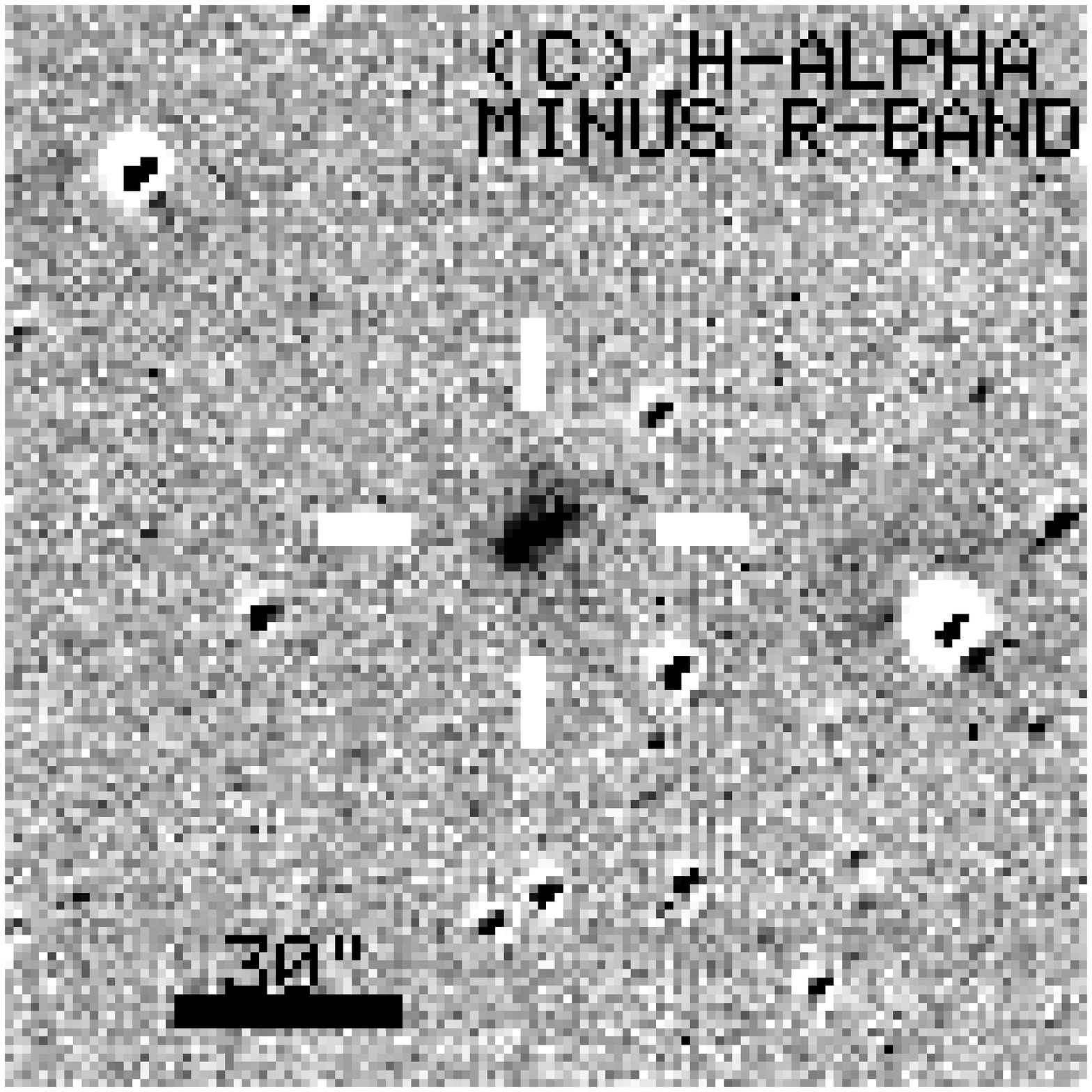}
\caption{\label{fig:images} The H~II region found in HIZSS~3.  In (a)
we show the H$\alpha$ image taken through a 30\AA\ wide interference
filter.  The H~II region is the extended feature in mid-frame.  
In (b) we show the continuum R-band exposure.  The H~II region is
faintly visible as the R bandpass includes the H$\alpha$ line. In (c) we show
the H$\alpha$ exposure with the continuum removed.  Differences in the 
point-spread-function lead to the over- and under-subtracted images for
the brighter stars.  In all three images,
north is up and east is the left.  The
fiducial mark in mid-frame denotes the center of the H~II region,
at $\alpha_{\rm J2000}=7^{\rm h}00^{\rm m}24.57^{\rm s}$ and 
$\delta_{\rm J2000}=-04^\circ13'13.7"$
}
\end{figure}

The H$\alpha$ source is located at 
$\alpha_{2000}= 7^{\rm h}00^{\rm m}24.57^{\rm s}$, 
$\delta_{2000}=-04^\circ13'13.7"$.  This
places it within a partially-resolved secondary peak in the H~I distribution
evident in 
Fig.~\ref{fig:totalHI}.
We've marked the location of the
H$\alpha$ source with an ``X" in Fig.~\ref{fig:totalHI}.  The H$\alpha$
source
is extended principally NW to SE, covering about 12 arcsecs.  At a
distance of 1.8~Mpc, this corresponds to a diameter of $\sim$100~pc, 
typical of
the ionization region (Str\"omgren sphere) of a late O-type star
(Spitzer 1968, with n$_e$=10 cm$^{-3}$) were it ionization-bounded.
More likely the region is density bounded (as is the case with an older,
evolved H~II region),
and the
size is 
consistent with the smallest
first-ranked H~II regions in irregular galaxies (see Table 2 of Youngblood \&
Hunter 1999).  Thus it may contain a modest population of OB stars.  The
bulge in the H~I contours suggest a partially-resolved H~I cloud,
and hence a likely home for star formation.

We were able to calibrate the H$\alpha$ image by observing several
spectrophotometric standards and by knowing the filter characteristics. 
The H$\alpha$ flux corresponds to $2.2\times 10^{-14}$ ergs cm$^{-2}$ s$^{-1}$.
If $E(B-V)=1.1$~mags, then $A_{\rm H\alpha}=2.8$~mags (Savage \& Mathis 1979),
and at a distance of 1.8~Mpc the H$\alpha$ luminosity would be 
$1.4\times 10^{38}$~ergs~s$^{-1}$.  This value is quite reasonable for
the luminosity of the brightest H~II regions within dwarf irregular
galaxies (Youngblood \& Hunter 1999), 
and is 14 times the H$\alpha$ luminosity of the
Orion nebula (Kennicutt 1984).
No other H$\alpha$ sources were found, either by blinking the
frames or by examining a continuum-subtracted version of the H$\alpha$ image.

The location and size of the H$\alpha$ source is certainly suggestive of
being part of
HIZSS~3.  However, we were concerned that this could instead
be a foreground planetary nebula.  To settle this matter, we obtained a
spectrum of the H$\alpha$ source using the KPNO 2.1-m telescope and GoldCam
spectrometer on 17 April 2003.  We used a 600 line mm$^{-1}$ grating (No.~35)
centered at $\lambda 5700$, giving us coverage from $\lambda 4500-7000$ at
1.24\AA\ pixel$^{-1}$.  A GG400 filter was used to block out any second-order
light. The slit of the spectrograph was rotated to a position
angle of 45$^\circ$, near the parallactic angle, and perpendicular to the
major extension.  The slit width was set to 100$\mu$m (1.3 arcsec) and provided 3.3\AA\ resolution.
The object was centered by a blind offset from a nearby star, and the exposure was 
hand-guided.
Conditions were marginal, 
and only one full 1200~s exposure was obtained before the cirrus thickened;
nevertheless, H$\alpha$ emission was strongly present at the predicted
location, with a signal level of 1500 e$^-$.
(A second exposure, terminated by clouds, showed emission at the same place.) 
A comparison He-Ne-Ar exposure was made both before and after the exposure.
The spectrophotometric standard star Feige 34 was observed (through clouds) to
provide {\it relative} flux calibration.  The nebular spectrum was
extracted using an exposure of the offset star to serve as a reference
for the 2-dimensional mapping of the spectrum on the chip.

\begin{figure}
\epsscale{1.1}
\plotone{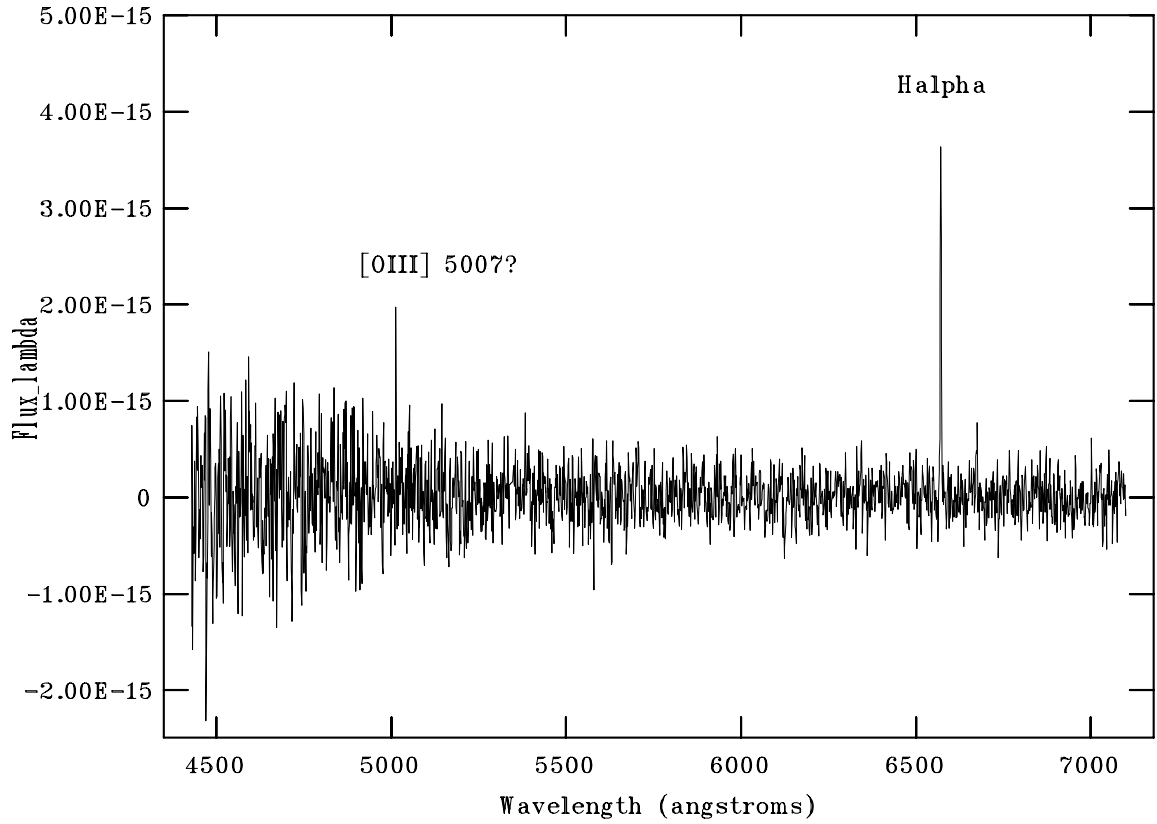}
\caption{\label{fig:spectrum} The spectrum of the H$\alpha$ source obtained
at the KPNO 2.1-m telescope.  Although the counts have been converted to
relative flux (in units of 
ergs~s$^{-1}$ cm$^{-2}$ \AA$^{-1}$), 
the non-photometric conditions and relatively narrow slit 
make the absolute scale somewhat unreliable; however, the integral across
the H$\alpha$ line produces an integrated flux that is only 1.5 times less
than that measured more accurately on our direct images.
}
\end{figure}

The spectrum is shown in Fig.~\ref{fig:spectrum}.  H$\alpha$ is strong, and
a weak feature may be [OIII] $\lambda 5007$.  The heliocentric radial
velocity of H$\alpha$ is measured to be 335$\pm$ 15 km~s$^{-1}$.  This is
in substantial agreement with the heliocentric 21-cm velocity map, which
suggests a velocity of 300 km~s$^{-1}$ at that position 
(cf.~Figs.~\ref{fig:totalHI} and \ref{fig:HIvel}), 
and confirms that the object we've identified in
our H$\alpha$ image is associated with HIZSS~3. The fact that
the values for the diameter and luminosities of the H~II region are quite
reasonable gives gives additional support to our interpretation. 

\section{Discussion}

Most recent discoveries of nearby galaxies have been of dwarf spheroid
systems which are not presently active in forming stars.  The presence
of substantial gas in HIZSS~3, as well as a modest-size H~II region,
suggests that this is a dwarf irregular (dIm).  As such, it is possibly
the nearest dIm to be discovered in the past 25 years; i.e., since
LGC~3 was found by Kowal, Lo, \& Sargent (1978),
and SagDIG was found by Cesarsky et al.~(1977) and Longmore et al.~(1978).
Both of these are well-established to be Local Group members. 
(For more on the growth of our knowledge of the Local Group, see
van den Bergh 2000.)

However, the 
properties of HIZSS~3 are most strongly reminiscent of those
of another nearby dIm, GR~8.  
This galaxy was first cataloged by Reaves (1956) in a survey of dwarf galaxies seen towards Virgo, but 
Hodge (1967) found it was a much nearer object, possibly even in the
Local Group.  GR8's optical appearance is dominated by 
a few bright H~II regions, the largest of which has a diameter of 175~pc
(Youngblood \& Hunter 1999), somewhat greater than the 100~pc found for
our H$\alpha$ source.  Carignan, Beaulieu, \& Freeman
(1990) find an H~I diameter for GR~8 of about 2~kpc, about two-thirds of
what we find for HIZSS~3. (We have adjusted these value to 
the 2.2~Mpc distance to GR~8
recommended by van den Bergh 2000, based upon newer distance measurements
by Tolstoy et al.\ 1995 and Dohm-Palmer et al.\ 1998 using a Cepheid and
the tip of the red giant branch, respectively.)

\begin{figure}
\epsscale{0.9}
\plotone{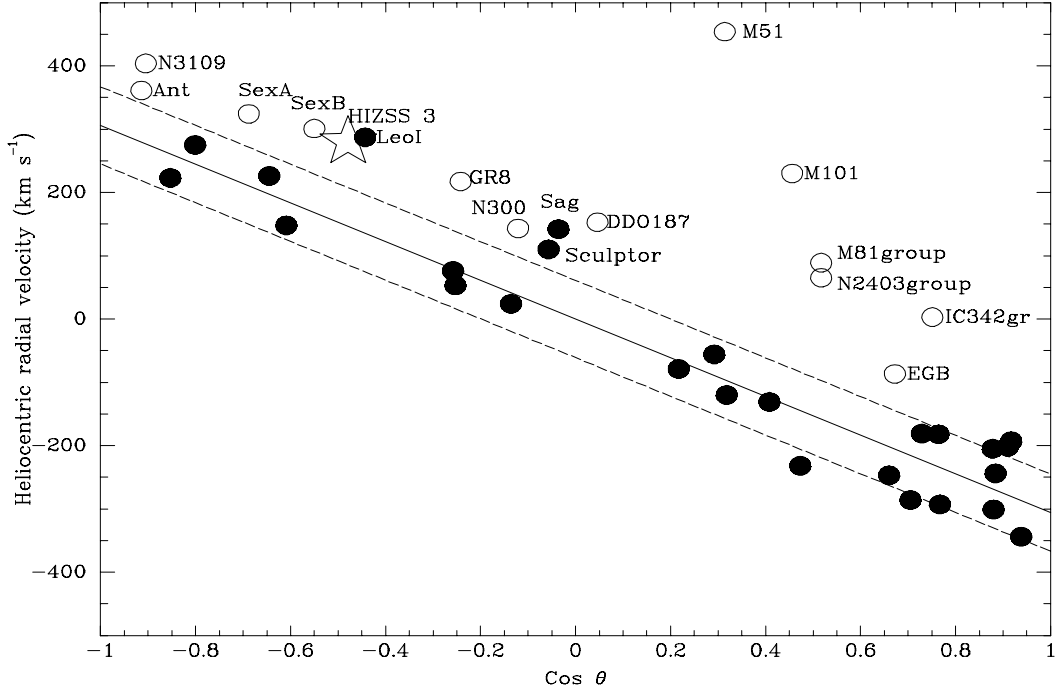}
\caption{\label{fig:sandage} The heliocentric radial velocity 
of the known 
Local Group members (solid points) follows a narrow band as a function of
$\cos \theta$, where $\theta$ is the angular distance of the galaxy from
the solar apex.  The exceptions are Leo~I, and the Sagittarius and
Sculptor dwarf spheroidals, the motions of which are strongly affected
by the Milky Way (van den Bergh 2000).
We have shown the relation found by Courteau \& van den Bergh
(1999) for Local Group galaxies by a solid line, and their $\pm 61$~km~s$^{-1}$ 1$\sigma$ envelope
by dashed lines.  
The open circles denote non-Local Group members.
The location of HIZSS~3 (denoted with a big star)
is similar to that of 
NGC 3109, the Antlia dwarf, and Sextans A, and Sextans B (non-members),
as well as that of Leo I (a satellite of the Milky Way).
The data come from Sandage (1986) and Courteau \& van den Bergh (1999) 
and references therein.}
\end{figure}

Could HIZSS~3 be a member of the Local Group?  We have adopted a tentative
distance to the galaxy of 1.8~Mpc based upon its radial velocity with
respect to the Local Group centroid (LGC) of 134~km~s$^{-1}$,  where
we have adopted the solution for the solar apex and solar motion (with
respect to the LGC) from Courteau \& van den Bergh (1999)\footnote{Note
that equation (6) in Courteau \& van den Bergh (1999) is incorrect, and
actually gives the correction to the LGC corresponding to the solution
of Yahil, Tammann, \& Sandage (1977).  The equation for the Courteau
\& van den Bergh (1999) solution should read: $V_{LGC}=V_{\rm helio}-48
\cos l \cos b +302 \sin l \cos b - 16 \sin b$.  This may be compared to
Sandage's (1986) solution using more limited data.  Sandage's solution
is expressed in terms of the motion of the LSR, but transformed to
the sun's motion corresponds to a velocity of 296 km s$^{-1}$ towards
$(l,b)=(95.0^\circ,-4.2^\circ)$.  Thus $V_{LGC}=V_{\rm helio}-26 \cos l
\cos b +296 \sin l \cos b - 22 \sin b$.  Adopting his solution for HIZSS~3
would lead to a velocity relative to the LGC of 120 km s$^{-1}$, and
hence a distance of 1.6~Mpc.} and used a value for the Hubble constant
$H_o=75$~km~s$^{-1}$~Mpc$^{-1}$.  Such a distance would place it well
beyond the zero-velocity boundary of the Local Group.  However, this
relies upon the assumption that the HIZSS~3 has no peculiar velocity.
In general, the velocities of nearby galaxies will be affected by the
presence of their neighbors.  However, the galaxies of the Local Group
do show only modest peculiar motions.  Yahil, Tammann, \& Sandage (1977),
Sandage (1986), and Courteau \& van den Bergh (1999), among others,  have
all found similar solutions for the velocities of the sun with respect
to the Local Group as a whole by minimizing the velocity residuals of
well-established Local Group members.  In Fig.~\ref{fig:sandage} we
show the heliocentric radial velocities of Local Group members (filled
circles) and that of nearby non-members (open circles) as a function
of $\cos \theta$, where $\theta$ is the angle from the solar apex.
The solution found by Courteau \& van den Bergh (1999) is shown by the
solid line, and corresponds to a motion of the sun of 306 km s$^{-1}$
towards $(l,b)=(99^\circ,-3^\circ)$.  The 1$\sigma$ scatter about this
solution is 61 km s$^{-1}$, shown by the dashed lines.  HIZSS~3 lies 
134~km~s$^{-1}$ 
above the solid line, i.e., 2.1$\sigma$ from the mean
relationship.  {\it Most} of the galaxies above the $+61$ (1$\sigma$)
dashed lines are non-members (NGC 3109, Sex A, Sex B, GR~8), but not
all: Leo~I, in particular, is in a very similar location in this diagram.
Measurements of the tip of the red giant branch (Lee et al.\ 1993) however
firmly establish the distance of this galaxy as 270~kpc, about one-tenth
one would infer from just its velocity. (The galaxy's peculiar velocity
is due to its proximity to the Milky Way.)  While we don't expect that
HIZSS~3 to be equally close, as that would be inconsistent with both
the properties of the H~II region and the 21-cm width relative to the
H~I flux,
its membership in the Local Group can not
be ruled out on the basis of the current data.
We conclude that HIZSS~3 is
unlikely to be a member of the Local Group based upon its velocity, but
emphasize the need for a better distance estimate.

Van den Bergh (2000) discusses galaxies on the outer fringes of the
Local Group, and argues that Sextans A, Sextans B, NGC 3109, and
Antlia form the nearest external group of galaxies beyond the Local Group
(see also van den Bergh 1999).  HIZSS~3 may be at a similar distance
and
does fall similarly above the LG relation in Fig.~\ref{fig:sandage}.
However, HIZSS~3 is located 46$^\circ$-49$^\circ$
from these galaxies, or about 1~Mpc away in linear space.  This is about
twice as far as the separation between Sextans A and NGC~3109.  We do not
know very much about the characteristics of such small groups, and it seems
conceivable, but unlikely, that HIZSS~3 is native to this group.
The identification
of other nearby galaxies in this region would be illuminating.

Followup observations of HIZSS~3 
are planned for the next observing season.  Of key
interest will be the detection of its resolved stellar population,
identification of other H~II regions (if any), 
and the determination of nebular chemical
abundances.  (Spectroscopy of the nebula potentially will lead to an 
improved reddening estimate.)
Measurements of an accurate distance would best be determined
in the infrared from the tip of the red giant branch, as
finding Cepheids (which vary most in the blue) will be greatly
hindered by foreground extinction.

\acknowledgments
We are thankful to Deidre Hunter for useful discussions, and for
helpful correspondence with Sidney van den Bergh.
We also thank Andrew Rivers for the VLA images, which originally
appeared in his Ph.~D. thesis.
The Kitt Peak
director, Richard Green, kindly allowed us to make use of the first hour
of an engineering night on the 2.1-m to obtain our optical spectrum, and
we gratefully acknowledge the assistance of Jim DeVeny and Ed Eastburn in
obtaining that observation. PM's role in this project was supported
by the National Science Foundation through grant AST0093060.

\begin{deluxetable}{l c }
\tabletypesize{\footnotesize}
\tablewidth{0pc}
\tablenum{1}
\tablecolumns{2}
\tablecaption{\label{tab:summary}Summary of Properties of HIZSS~3}
\tablehead{
\colhead{Property}
&\colhead{Value}
}
\startdata
$\alpha_{\rm J2000},\delta_{\rm 2000}$ (H~I peak) & $7^{\rm h}00^{\rm m}29.3^{\rm s},
-04^\circ12'30"$ \\
Galactic coords.~$l,b$& 217.71$^\circ$, $+0.09^\circ$ \\
Heliocentric radial velocity (H~I peak) & 280 km s$^{-1}$ \\
Radial velocity wrt Local Group centroid & 134 km s$^{-1}$ \\
Inferred distance ($H_o=75$ km s$^{-1}$ Mpc$^{-1}$) & 1.8~Mpc \\
H~I velocity width (20\% peak intensity), single-dish (Parkes) & 85 km s$^{-1}$ \\ 
H~I velocity width (20\% peak intensity), VLA & 91 km s$^{-1}$ \\ 
H~I diameter (angular) & 6' \\
H~I diameter (linear, for 1.8~Mpc distance) &  3~kpc \\
Integrated 21-cm flux-density, single-dish (Parkes) & 32.1 Jy km s$^{-1}$\\
H~I mass (for 1.8~Mpc distance),single-dish (Parkes) & $2.5\times 10^{7}$ M$_\odot$ \\
Estimated foreground $E(B-V)$ &  1.1~mags \\
$\alpha_{\rm J2000},\delta_{\rm 2000}$ (HII region) & 
$7^{\rm h}00^{\rm m}24.57^{\rm s},
-04^\circ13'13.7"$ \\
H~II region diameter (angular)& 12" \\
H~II region diameter (linear, for 1.8~Mpc distance) & 100~pc \\
H~II region H$\alpha$ flux & $2.2\times 10^{-14}$ ergs cm$^{-2}$ s$^{-1}$ \\
H~II region H$\alpha$ luminosity  (for 1.8~Mpc distance)& $1.4\times 10^{38}$~ergs~s$^{-1}$ \\
\enddata
\end{deluxetable}
\clearpage

\end{document}